\documentclass[a4paper,fleqn,usenatbib,useAMS]{mnras}

\usepackage{graphicx}   
\usepackage{color}
\usepackage{lscape}
\usepackage{txfonts}
\usepackage{amsfonts}

\title[$\Gamma_{\rm 2-10keV}-L_{\rm bol}/L_{\rm Edd}$ correlation]
{A systematic study of the condensation of the corona and the application for 
$\Gamma_{\rm 2-10keV}-L_{\rm bol}/L_{\rm Edd}$ correlation in luminous active galactic nuclei}

\author[Erlin Qiao and B.F. Liu]{Erlin Qiao $^{1,2}$\thanks{E-mail:
qiaoel@nao.cas.cn} and B.F. Liu $^{1,2}$\\
$^{1}$Key Laboratory of Space Astronomy and Technology, National Astronomical Observatories, Chinese Academy of
Sciences, Beijing 100012, China \\
$^{2}$School of Astronomy and
Space Sciences, University of Chinese Academy of Sciences, 19A Yuquan Road, Beijing 100049, China\\}

\date{Accepted XXX. Received YYY; in original form ZZZ}

\pubyear{2017}

\begin{document}
\label{firstpage}
\pagerange{\pageref{firstpage}--\pageref{lastpage}}
\maketitle
\begin{abstract}
In this paper, we explained  the observed $\Gamma_{\rm 2-10keV}-L_{\rm bol}/L_{\rm Edd}$ correlation 
in luminous active galactic nuclei within the framework of the condensation of the corona around
a supermassive black hole \citep{Liu2015,Qiao2017}. 
Specifically, we systemically test the effects of black hole mass $M$, the viscosity parameter $\alpha$, and the 
magnetic parameter $\beta$ (with magnetic pressure $p_{\rm m}=B^2/{8\pi}=(1-\beta)p_{\rm tot}$, $p_{\rm
tot}=p_{\rm gas}+p_{\rm m}$) on the structure of the accretion disc and the corona, as well as the corresponding 
emergent spectra. It is found that the hard X-ray photon index $\Gamma_{\rm 2-10keV}$ nearly does not change with changing 
black hole mass $M$, or changing magnetic parameter $\beta$.
Meanwhile, it is found that the geometry of the accretion flow, i.e., the relative configuration of the 
disc and corona,  as well as the emergent spectra can be strongly affected by changing the value of $\alpha$.
By comparing with a sample composed of 29 luminous active galactic nuclei
with  well constrained X-ray spectra and Eddington ratios, 
it is found that the observed  $\Gamma_{\rm 2-10keV}-L_{\rm bol}/L_{\rm Edd}$ correlation 
can be well matched with a relatively bigger value of $\alpha$, i.e., $\alpha \sim 1$, as previously also suggested 
by \citet[]{Narayan1996} for luminous accreting black holes.  
\end{abstract}


\section{Introduction}
The multi-band observations for the luminous active galactic nuclei (AGNs) 
support that the accretion flow around the central supermassive black hole exists
in the  form  of a  disc-corona structure \citep[e.g.][]{Haardt1991,Haardt1993,
Svensson1994,Stern1995}.
The optical/ultraviolet (UV) emission is often explained to be from the cool disc
extending down to the innermost stable circular orbits (ISCO) of the black hole, while X-ray emission is 
often explained to be from the inverse Compton scattering of the  
optical/UV photons from the disc in the hot corona above
\citep[e.g.][]{Magdziarz1998,Gierlinski2004,Piconcelli2005,Vasudevan2007,Vasudevan2009,Fabian2015,Lusso2017}. 
In such a scenario, a long-standing problem is that how the hot corona is formed and heated.
In the previous disc-corona models, it was often needed to assume a large fraction of the 
accretion energy to be released in the corona to explain the observed strong X-ray emissions in luminous     
AGNs, including the bright Seyfert galaxies and quasars 
\citep[e.g.][]{Haardt1991,Haardt1993,Svensson1994,Stern1995,Dove1997}. 
Currently, a promising mechanism for the heating of the corona 
is the magnetic reconnection \citep[e.g.][]{DiMatteo1998,Merloni2001,Liu2002b,Liu2003,Kawanaka2005,
Uzdensky2008,Cao2009}.
The electron distribution in the corona is very likely to  have a non-thermal tail (i.e., to be hybrid) 
due to the magnetic reconnection \citep[][]{Poutanen1998,Coppi1999,Poutanen2014,Poutanen2017}.
By means of large-scale particle-in-cell (PIC) simulations in both 2D and 3D, \citet[]{ Sironi2014} 
believed that they have provided a definitive evidence for the  
non-thermal particle acceleration via magnetic reconnection, which has very important implications for 
understanding the generation of the hard particle spectra in relativistic astrophysical plasmas. 
So far, a great of very important progresses have been achieved to establish the disc and the corona via  
magnetohydrodynamic (MHD) simulations \citep[e.g.][]{Miller2000,Hirose2006,
Loureiro2012,Bai2013,Fromang2013,Uzdensky2013,Takahashi2016}.
However, it is still difficult to compare the  results of 
MHD simulations with the observed spectra. Some papers, such as  \citet{Jiang2014} estimated 
the fraction of the accretion energy dissipated in the coronal region above the accretion disc based on 
the three-dimensional radiation MHD simulations. They found that  the maximum fraction of the 
accretion energy dissipated in the corona is only 3.4\%, which can not explain the strong X-ray emissions
observed in luminous AGNs \citep[e.g.][]{Elvis1994,Shang2005,Vasudevan2007,Vasudevan2009}. 

In order to explain the origin of the strong X-ray emissions in luminous AGNs, and avoid the 
complicated processes of the  magnetic reconnection,
a hot accretion model was proposed by
\citet{Liu2015}.  In this model,  it is assumed that, initially, beyond the Bondi radius,
a vertically extended hot coronal  gas is supplied to the central black hole by capturing the
interstellar medium and the hot stellar wind. In \citet{Liu2015}, the hot corona is 
described as the self-similar solution  of the advection-dominated accretion flows (ADAF) \citep[][]{Narayan1994} . 
The ADAF is a two-temperature hot accretion flow, in which the ions are directly heated by 
the general viscous heating, and the electrons are heated by 
the ions via Coulomb collision  \citep[][]{Narayan1995} . 
Since  the angular velocity of the  ADAF solution  is intrinsically  sub-Keplerian, the viscous 
time scale is much shorter than that of the cooling time scale in  ADAF solution \citep[][for review]{Narayan2008,Yuan2014}.
In this case, generally the hot coronal gas will be accreted directly towards the central black hole instead of
collapsing locally immediately.  Specifically, there exists a critical mass accretion rate $\dot M_{\rm crit}$.
For $\dot M \gtrsim \dot M_{\rm crit}$, the  hot coronal gas flows towards the black hole until at a critical radius
$r_{\rm d}$ (in units of Schwarzschild radius $R_{\rm S}$, with $R_{\rm S}=2.95\times 10^5 M/M_{\odot} \ \rm {cm}$), 
a fraction of the hot  coronal gas begins to condense to the equatorial plane of the black hole, forming an inner
disc. Then these gases will be accreted in the form of  a disc-corona structure extending down to the ISCO of the black hole, 
maintaining a relatively stronger X-ray emissions.
For $\dot M \lesssim \dot M_{\rm crit}$, the gases will be accreted in the form of a pure ADAF without condensation.
The size of the inner disc increases with increasing $\dot M$. For clarity, one can refer to 
Fig. 1 in \citet{Qiao2017} for the schematic description of the model.
The model of the  condensation of the hot corona  predicts a very different geometry of the accretion flow,
which  preliminarily has been employed to explain the observed correlation between the 
X-ray slope and the Compton reflection scaling factor in luminous AGNs \citep{Qiao2017}.
Meanwhile, it is shown that, although geometrically the size of the  hot corona  is very extended in the radial direction,
more than $80\%$ of the  hard X-rays are emitted in a very small region 
less than $\sim 10$ Schwarzschild radii from the view point of radiation,  
which is consistent with the X-ray mapping observations for estimating the size of 
the corona in luminous AGNs \citep[e.g.][]{Liu2017,Reis2013,Fabian2015}.

Observationally, it is found that  there is an anti-correlation between the hard X-ray photon index $\Gamma_{\rm 2-10keV}$
and Eddington ratio $L_{\rm bol}/L_{\rm Edd}$ (with $L_{\rm Edd}=1.26\times 10^{38} M/M_{\odot}\ \rm {erg s^{-1}}$) 
for $L_{\rm bol}/L_{\rm Edd} \lesssim 0.01$ in low-luminosity AGNs, 
which is simply interpreted in the framework of ADAF solution \citep[e.g.][]{Gu2009,Younes2011,
Emmanoulopoulos2012,Hernandez2013,Hernandez2014}.
While separately it is found that there is a positive correlation between  $\Gamma_{\rm 2-10keV}$ and $L_{\rm bol}/L_{\rm Edd}$ for 
$L_{\rm bol}/L_{\rm Edd} \gtrsim 0.01$ \footnote{Please note that in different literatures the authors use different
methods to estimate the bolometric luminosity $L_{\rm bol}$, which can have obvious effect on the slope of
the $\Gamma_{\rm 2-10keV}-L_{\rm bol}/L_{\rm Edd}$ correlation.
One can refer to the references we list for details.} in luminous AGNs, 
\citep[e.g.][]{Lu1999,Porquet2004,Wang2004,Shemmer2006,Saez2008,Sobolewska2009,Veledina2011}
which is often interpreted in the framework of the disc-corona model with the corona 
heated by magnetic reconnection \citep[e.g.][]{Cao2009,Liu2016}.
We suggested that both the anti-correlation between $\Gamma_{\rm 2-10keV}$ and  $L_{\rm bol}/L_{\rm Edd}$ 
in low-luminosity AGNs and the positive correlation between $\Gamma_{\rm 2-10keV}$ and 
 $L_{\rm bol}/L_{\rm Edd}$ in luminous AGNs can be unified to be described 
within the framework of the condensation of the hot corona/ADAF. 
Based on our model, when $\dot M \lesssim \dot M_{\rm crit}$,  the condensation does not occur,
the gas is accreted in the form of the pure ADAF. In the regime of ADAF solution 
with $10^{-4} \dot M_{\rm Edd}  \lesssim  \dot M \lesssim \dot M_{\rm crit}$ 
(with $\dot M_{\rm Edd}$ = $1.39 \times 10^{18} M/M_{\rm \odot} \rm \ g s^{-1}$), the
X-ray emission is dominated by the inverse Compton scattering of the  
synchrotron radiation  and  bremsstrahlung photons  of the ADAF itself. 
In this case, the Compton scattering optical depth increases with increasing $\dot M$, while the 
electron temperature in the ADAF is nearly a constant, resulting in an increase of
the Compton parameter $y$ with increasing $\dot M$ \citep[e.g.][]{Mahadevan1997}. Then the X-ray 
spectra will become harder with increasing $\dot M$. More specifically, an anti-correlation between 
$\Gamma_{\rm 2-10keV}$ and  $L_{\rm bol}/L_{\rm Edd}$
will be predicted in the regime of ADAF solution \citep[e.g.][]{Qiaoetal2013,Qiao2013,Yang2015}.

In this paper, we will focus on the positive correlation between $\Gamma_{\rm 2-10keV}$ and 
$L_{\rm bol}/L_{\rm Edd}$ in luminous AGNs based on the model of the condensation of the corona/ADAF 
\citep[][]{Liu2015,Qiao2017}.
In the condensation model, an inner disc 
forms for $\dot M \gtrsim \dot M_{\rm crit}$. Both the outer boundary of the inner disc and the 
condensation rate of the corona increase with increasing $\dot M$. In this case, 
with an increase of $\dot M$, more soft photons from the disc will be scattered in the corona, 
which will result in a lower electron temperature in the corona, consequently predicting a softer X-ray spectrum.
Specifically, a positive correlation between $\Gamma_{\rm 2-10keV}$ and  $L_{\rm bol}/L_{\rm Edd}$
will be predicted for $\dot M \gtrsim \dot M_{\rm crit}$.
We systemically studied the effects of the  parameters in the model, such as the black hole $M$, the 
viscosity parameter $\alpha$, and the magnetic parameter $\beta$  
on the structure of the disc and corona, as well as the corresponding emergent spectra
around a supermassive black hole in AGNs.    
We compared the theoretical relation between $\Gamma_{\rm 2-10keV}$ and $L_{\rm bol}/L_{\rm Edd}$ with 
the observations of a sample composed of 29 luminous AGNs \citep{Vasudevan2009}. 
It is found that a bigger value of $\alpha \sim 1$ is required to 
match the observation, which is intrinsically  consistent with the nature of the ADAF solution applied to 
luminous accreting black holes.
The model is briefly introduced in Section 2. The numerical results and the comparisons with observations 
are shown in Section 3. Some discussions are in Section 4, and the conclusions are in Section 5.

\section{The model}
We consider that, initially, a vertically extended hot coronal gas is supplied 
to the central supermassive black by capturing the interstellar medium and stellar wind. 
Such a hot coronal gas is described as the self-similar solution of ADAF \citep[][]{Narayan1995}.
We study the interaction of the corona with a pre-existing, optically thick, cold disc 
in the  vertical direction. The interaction between the disc and the corona
results in either the matter in the corona to be condensed into the disc or the matter in the disc 
to be evaporated into the corona. 
Specifically, we study the energy balance including the thermal conduction in the vertical direction
between the disc and corona, the radiation of the corona (involving the  
synchrotron radiation, bremsstrahlung and the corresponding 
self-Compton scattering of the synchrotron radiation and bremsstrahlung photons in the corona 
itself, as well as the soft photons from the disc to be scattered in the corona), 
and the illumination of the disc by the corona
to self-consistently calculate the coronal 
temperature, density, and the mass-exchange rate between the disc and the corona. 
One can refer to \citet{Liu2015} or \citet{Qiao2017}  for the detailed model descriptions.
The structure of the disc and corona can be derived by specifying 
the black hole mass $m$ ($m=M/M_{\odot}$), initial mass accretion rate $\dot m$ 
($\dot m=\dot M/\dot M_{\rm Edd}$), the viscosity parameter $\alpha$,
magnetic parameter $\beta$ and the albedo $a$ (defined as the ratio of the reflected radiation from 
the surface of the inner disc to the incident radiation on it from the corona).
Numerical simulations show that the albedo is often very lower, i.e., $a \sim 0.1-0.2$ \citep{Magdziarz1995,
Zdziarski1999,Nayakshin2000}. In all the calculations below, we fix $a=0.15$.
With the derived structure of the disc and the corona, including 
the distribution of the electron temperature $T_{\rm e}$, the scattering optical depth $\tau_{\rm es}$ 
of the corona, and the surface effective temperature of the disc in the radial direction,  
we calculate the emergent spectrum of the disc-corona system
with Monte Carlo simulations. One can refer to \citet[][]{Qiao2012} for details.

\section{Numerical results}
\subsection{The effect of black hole mass}\label{secmass}
\begin{figure*}
\includegraphics[width=88mm,height=65mm,angle=0.0]{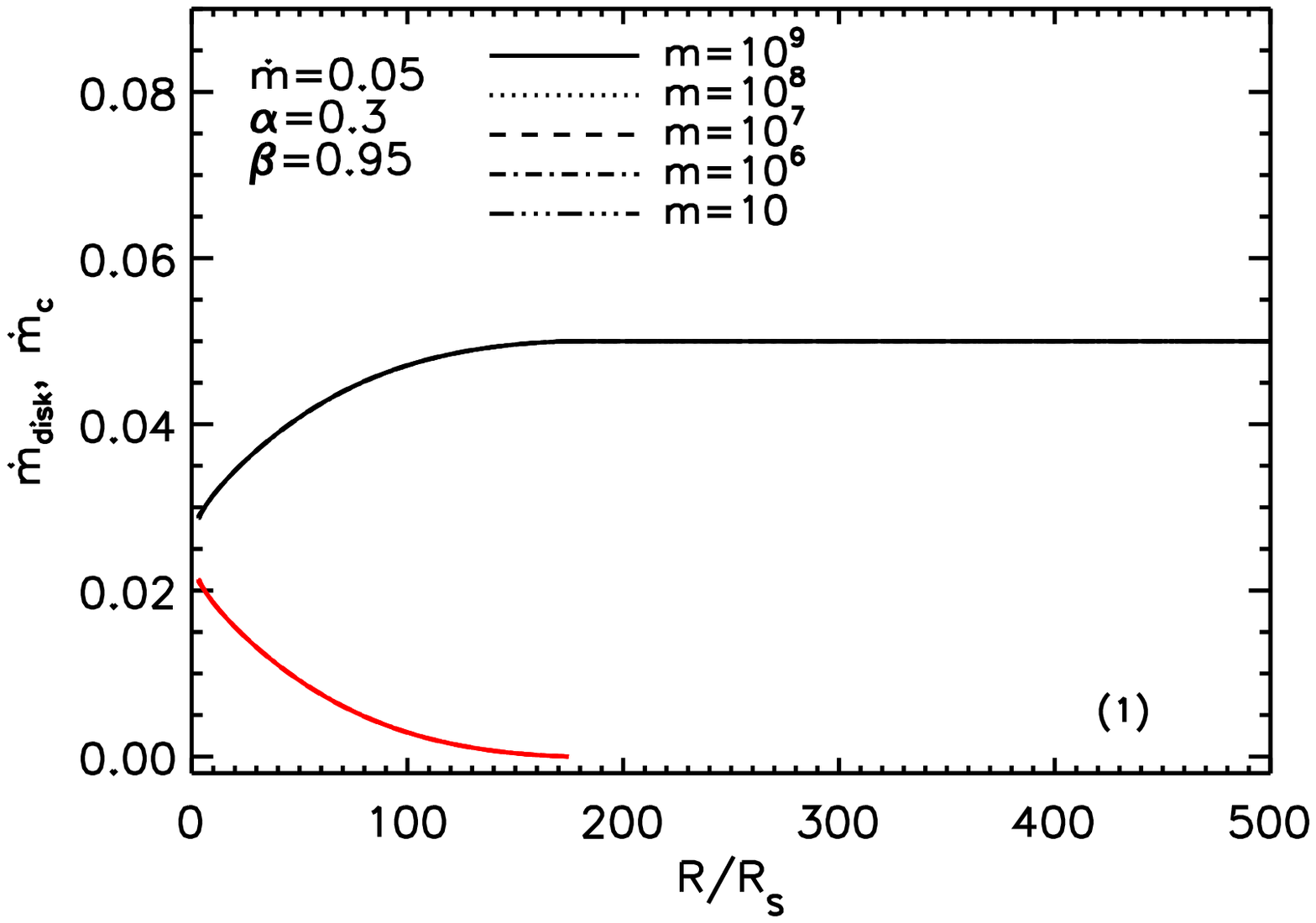}
\includegraphics[width=88mm,height=65mm,angle=0.0]{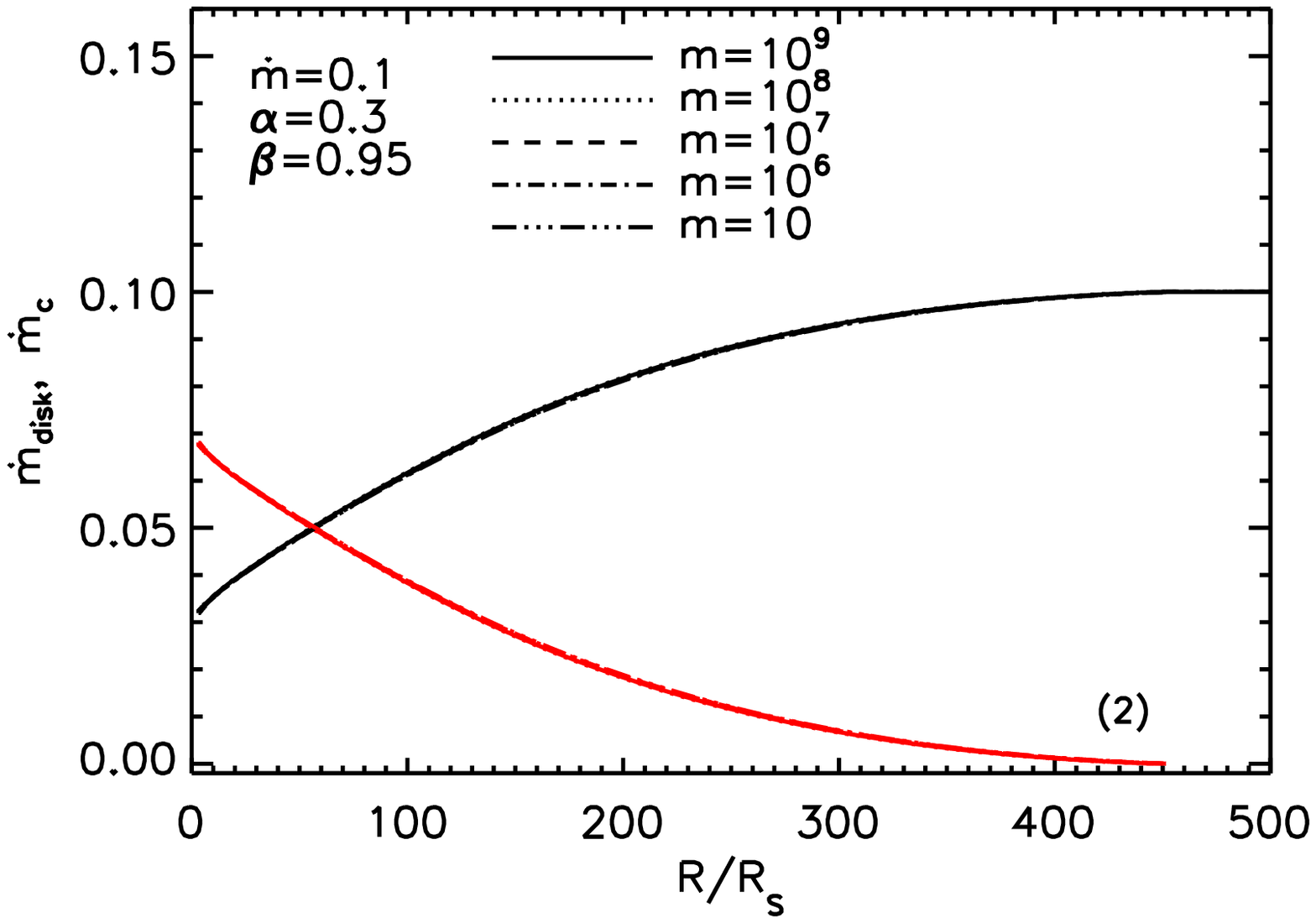}
\includegraphics[width=88mm,height=65mm,angle=0.0]{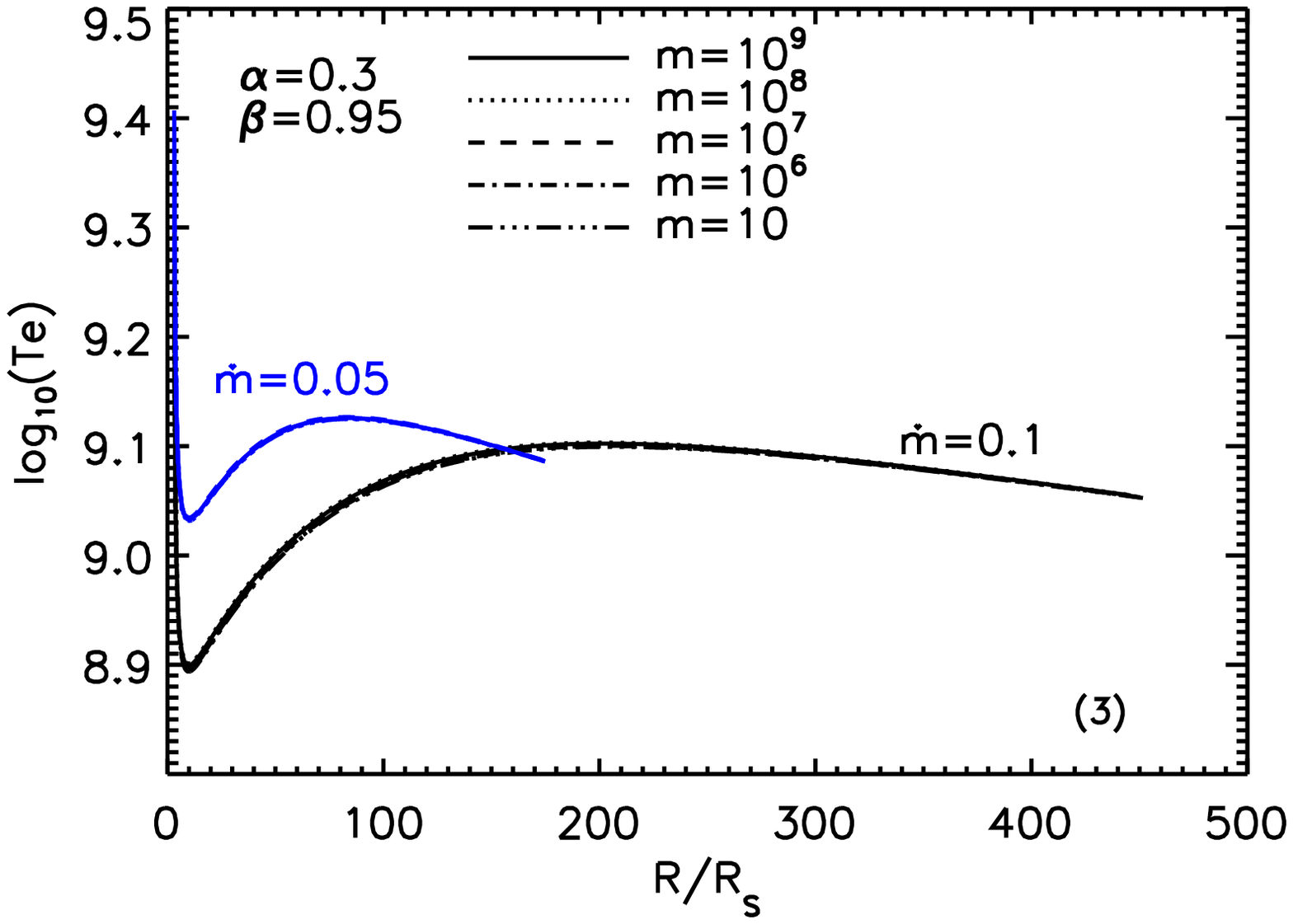}
\includegraphics[width=88mm,height=65mm,angle=0.0]{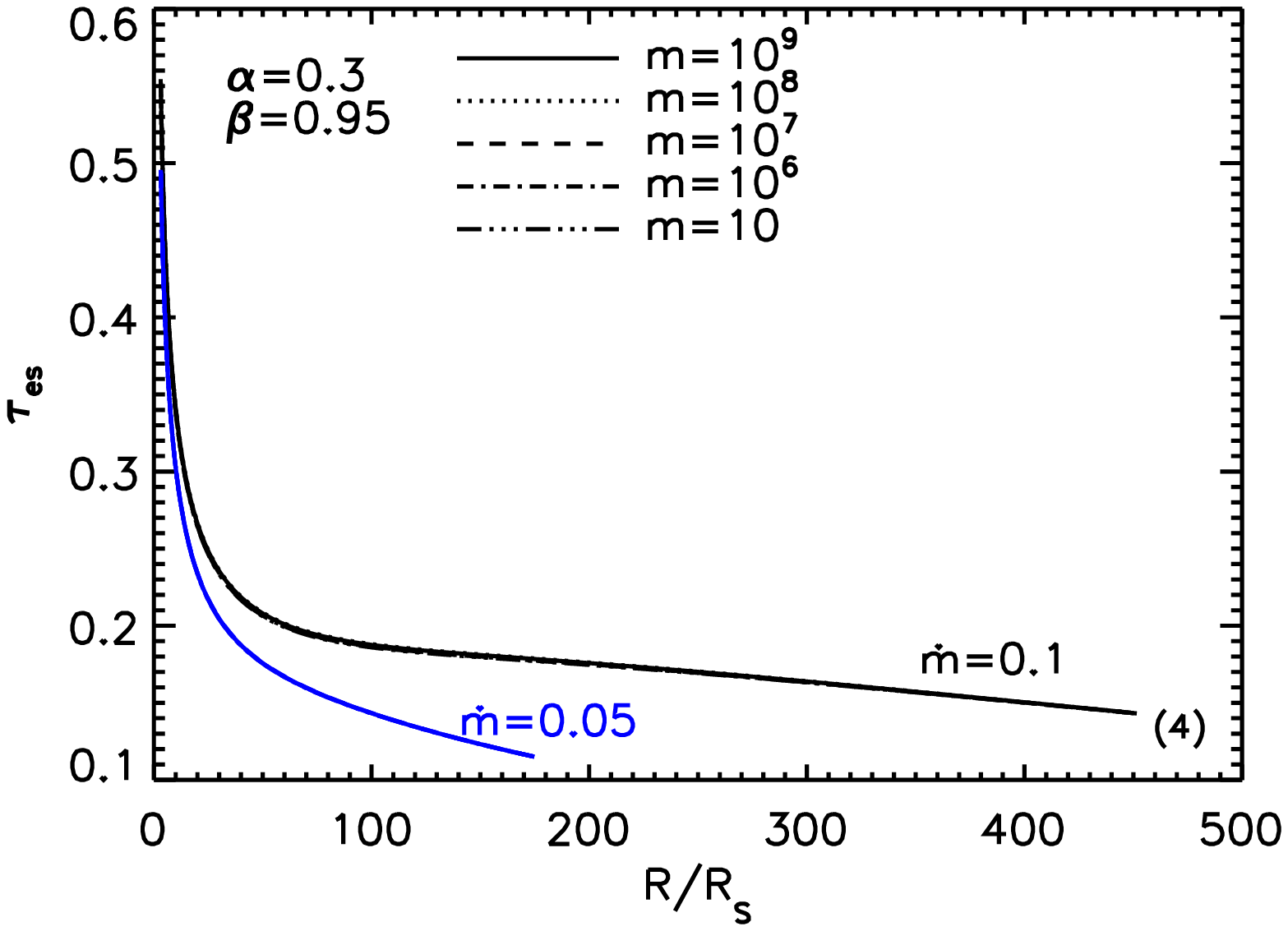}
\caption{\label{mass1}}Panel (1): mass accretion rate in the corona (black line) and mass
accretion rate in the disc (red line)  as a function of radius  with  the initial mass accretion rate $\dot m=0.05$
for different black hole masses. 
Panel (2): mass accretion rate in the corona (black line) and mass
accretion rate in the disc (red line)  as a function of radius  with  the initial mass accretion rate $\dot m=0.1$
for different black hole masses. 
Panel (3): electron temperature $T_{\rm e}$  in the corona as a function of radius with the initial mass accretion rate 
$\dot m=0.05$ (blue line) and $\dot m=0.1$ (black line) for different black hole masses. 
Panel (4): Compton scattering optical depth $\tau_{\rm es}$  in the corona as a function of radius with 
the initial mass accretion rate $\dot m=0.05$ (blue line) and $\dot m=0.1$ (black line) for different black hole masses. 
\end{figure*}

\begin{figure*}
\includegraphics[width=88mm,height=65mm,angle=0.0]{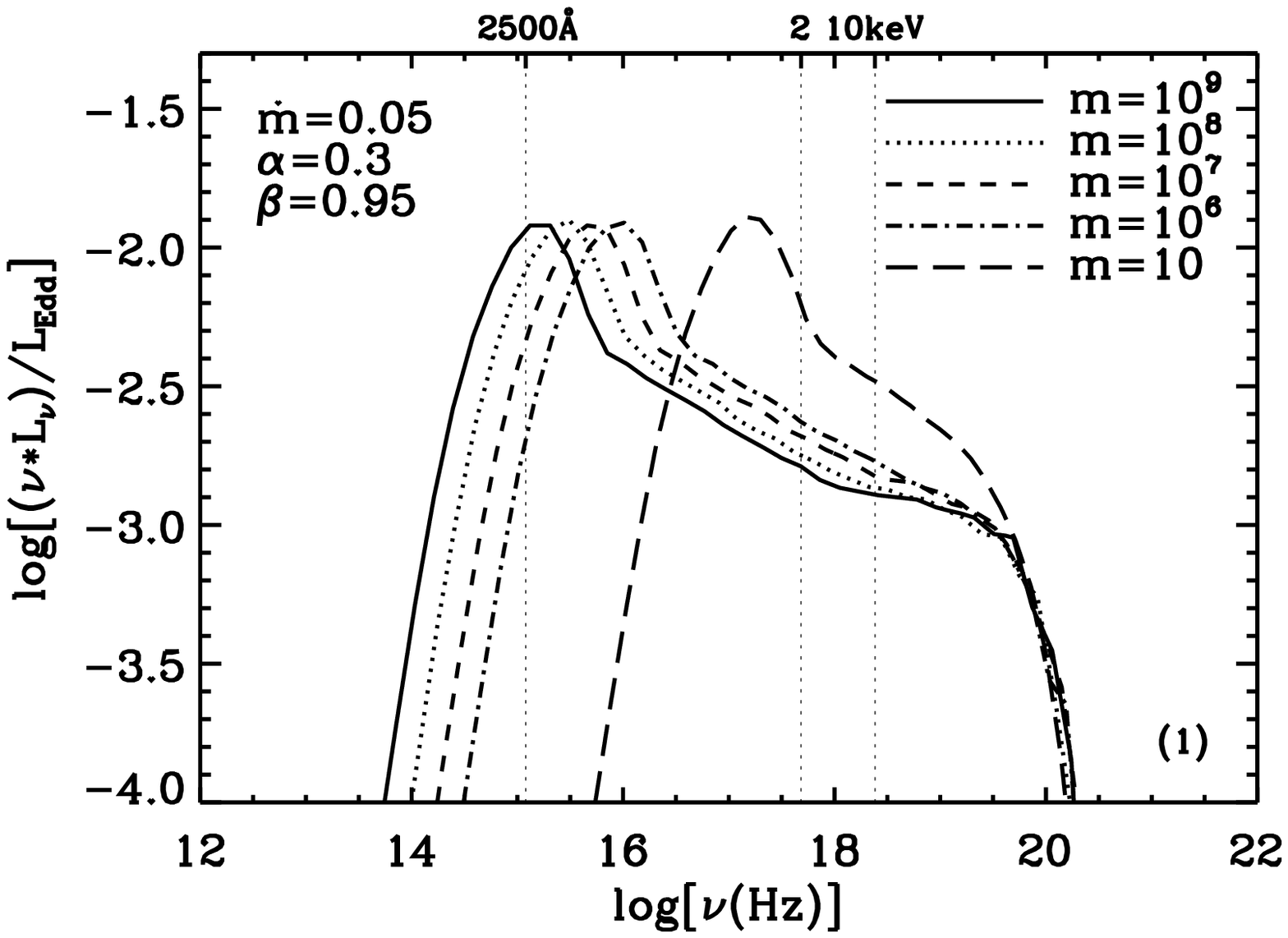}
\includegraphics[width=88mm,height=65mm,angle=0.0]{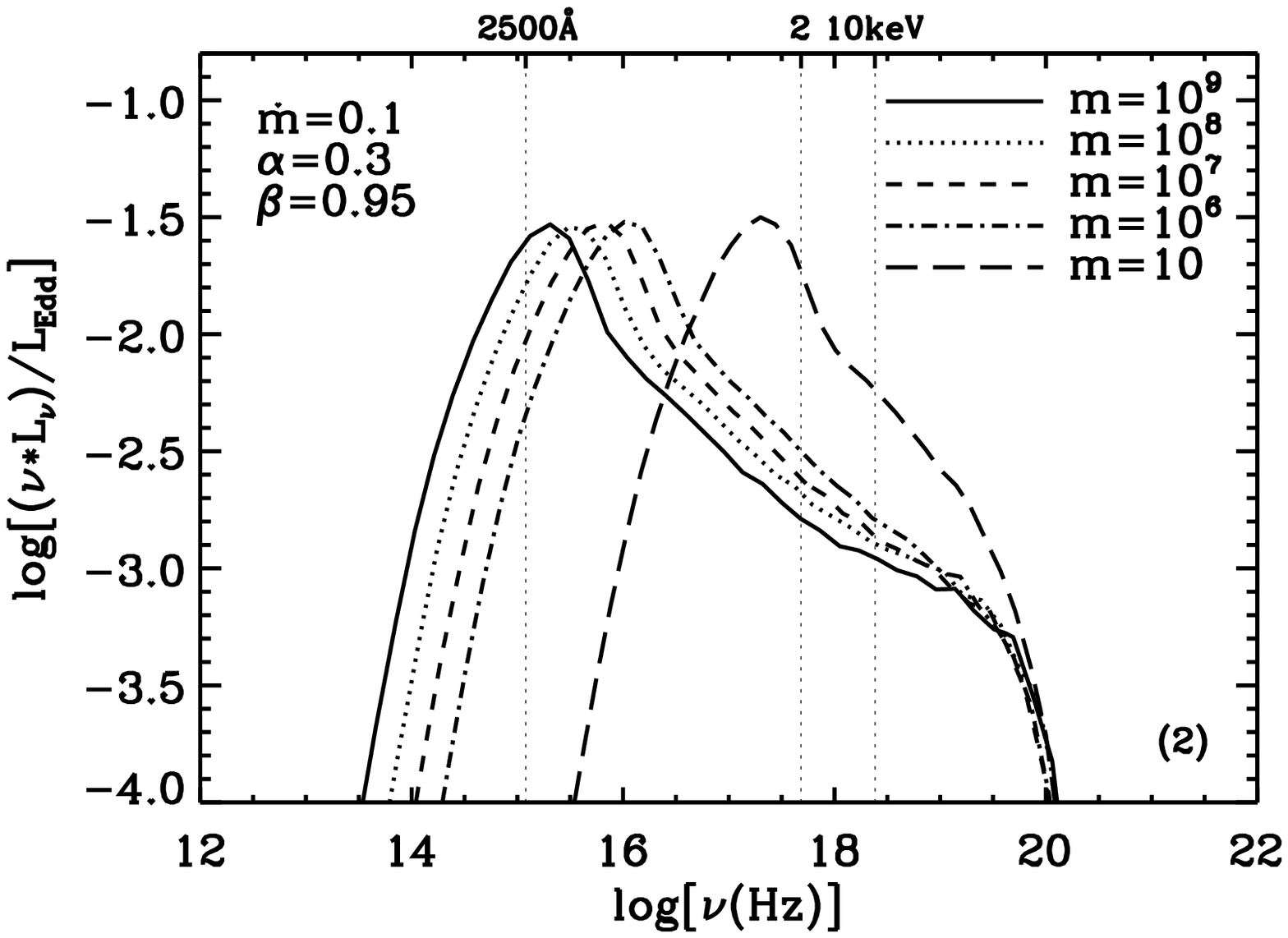}
\includegraphics[width=88mm,height=65mm,angle=0.0]{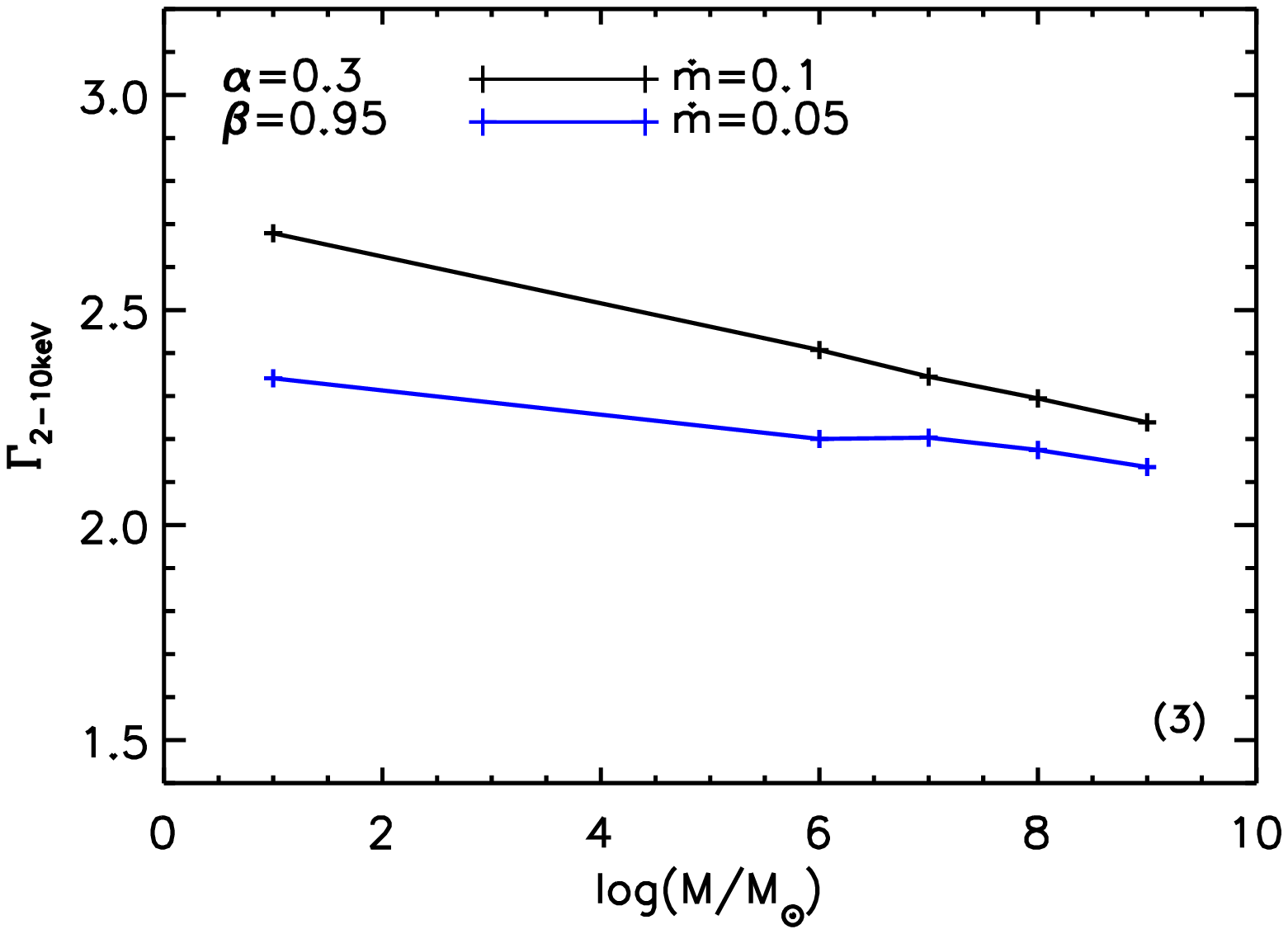}
\includegraphics[width=88mm,height=65mm,angle=0.0]{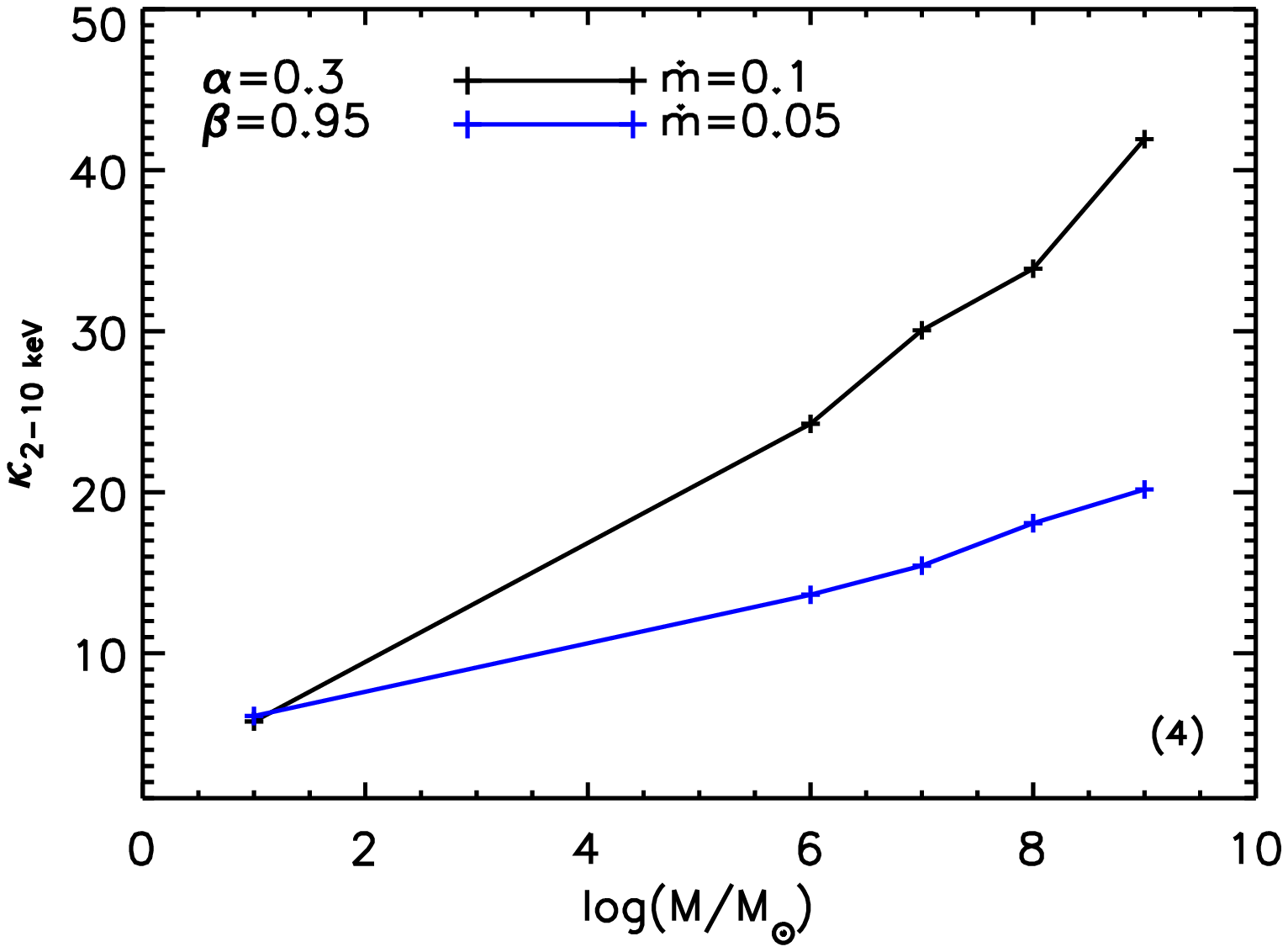}
\caption{\label{mass2}} Panel (1): emergent spectra with the initial mass accretion rate $\dot m=0.05$ 
for different black hole masses. 
Panel (2): emergent spectra with the initial mass accretion rate $\dot m=0.1$ 
for different black hole masses. 
Panel (3): hard X-ray photon index $\Gamma_{\rm 2-10keV}$ as a function of black hole mass $m$ for $\dot m=0.05$ (blue line) 
and $\dot m=0.1$ (black line).
Panel (4): $\kappa_{\rm 2-10keV}$ as a function of black hole mass $m$ for $\dot m=0.05$ (blue line) 
and $\dot m=0.1$ (black line).
\end{figure*}

In the panel (1) and panel (2) of Fig. \ref{mass1}, we plot the mass accretion rate in the corona (black line) and mass 
accretion rate in the disc (red line)  as a function of radius  with initial mass accretion rate 
$\dot m=0.05$ and  $\dot m=0.1$ for black hole
mass $m=10^9, 10^8, 10^7, 10^6$ and $10$ respectively. It is found that the effects of the black hole 
mass on the distribution of the mass accretion rate in either corona or disc  nearly can  be neglected.  
The condensation radius  $r_{\rm d}$
and the condensation rate of the corona $\dot m_{\rm cnd}$
(with $\dot m_{\rm cnd}$ being the integrated condensation rate from the condensation radius  
$r_{\rm d}$ to the ISCO of the black hole in units of $\dot M_{\rm Edd}$. With ISCO being 3$R_{\rm S}$ assumed for 
a non-rotating Schwarzschild black hole) are also very similar for different black hole masses with 
a fixed initial mass accretion rate.  One can refer to Table \ref{mass_effect} for details.
In the  panel (3) and panel (4) of Fig. \ref{mass1}, we plot the electron temperature  $T_{\rm e}$ and the 
Compton scattering optical depth $\tau_{\rm es}$ of the corona  as a function of radius  with mass accretion rate 
$\dot m=0.05$ (blue line)  and $\dot m=0.1$ (black line)  for black hole
mass $m=10^9, 10^8, 10^7, 10^6$ and $10$ respectively. It is also clear that the effects of the black hole mass
on the  distribution of the electron temperature and the Compton scattering optical depth nearly can be neglected.   
In the panel (1) and panel (2) of Fig. \ref{mass2}, we plot the corresponding emergent spectra (in units of $L_{\rm Edd}$)
with mass accretion rate $\dot m=0.05$ and  $\dot m=0.1$ for black hole
mass $m=10^9, 10^8, 10^7, 10^6$ and $10$ respectively.  The maximum effective temperatures of the disc
are $T_{\rm eff, max}=3.2, 5.7, 10.0, 18.0$ and $320.5\ {\rm eV}$   
with $\dot m=0.05$ for black hole mass $m=10^9, 10^8, 10^7, 10^6$ and $10$ respectively, and 
$T_{\rm eff, max}=4.0, 7.1, 12.7, 22.6$ and $402.5\ {\rm eV}$   
with $\dot m=0.1$ for black hole mass $m=10^9, 10^8, 10^7, 10^6$ and $10$ respectively.
As we expect, the emergent spectrum shifts rightward with decreasing $m$. 
We plot the hard X-ray photon index $\Gamma_{\rm 2-10keV}$  as a function of $m$
for $\dot m=0.05$ (blue line) and  $\dot m=0.1$ (black line) in the panel (3) of Fig. \ref{mass2}.  
It is shown that $\Gamma_{\rm 2-10keV}$ nearly does not change (a slight decrease) with increasing $m$, 
which can be understood as follows.
In our model, the power-law hard X-ray emission is produced by the inverse Compton scattering of the photons 
from the disc and corona itself (synchrotron radiation and bremsstrahlung) in the corona. As we know, 
the hard X-ray index is determined by the combination of $T_{\rm e}$ and $\tau_{\rm es}$. 
Specifically, for the optically thin case of the corona, $\Gamma$ can be simply expressed as,
\begin{eqnarray}\label{index}
\Gamma=1-{\ln{\tau_{\rm es}}\over {\ln{A}}},
\end{eqnarray} 
where $A=1+4\theta_{\rm e}+16 {\theta_{\rm e}}^2$ ($\theta= {kT_{\rm e}\over {m_{\rm e}c^2}}$,
with $k$ being the  Boltzmann constant, $m_{\rm e}$ being the mass of electron and $c$ being the speed of light) 
is the mean amplification factor in one scattering. As we show 
in the panel (3) and panel (4) of Fig. \ref{mass1}, $T_{\rm e}$ and $\tau_{\rm es}$ nearly do not change with
increasing $m$, consequently $\Gamma$ nearly does not change with increasing $m$.
By integrating the emergent spectra, we plot the correction factor $\kappa_{\rm 2-10keV}$ 
(with $\kappa_{\rm 2-10keV}=L_{\rm bol}/L_{\rm 2-10keV}$) 
as a function of $m$ for $\dot m=0.05$ (blue line) and  $\dot m=0.1$ (black line) in the  panel (4) of Fig. \ref{mass2}.
Here, we show that $\kappa_{\rm 2-10keV}$ intrinsically increases with increasing $m$ for a fixed mass
accretion rate, which we expect to be compared with observations in detail in the future.

\begin{table*}
\caption{Condensation and spectral features of the inner disc and corona for different black hole masses.
$r_{\rm d}$ is the condensation radius. $\dot m_{\rm cnd}$ is the integrated condensation rate.
$T_{\rm eff,max}$ is the maximum effective temperature of the inner disc.  
$\Gamma_{\rm 2-10 keV}$ is the hard X-ray photon index between 2 and 10keV.} 

\centering
\begin{tabular}{ccccccccc}
\hline\hline
$m$ & $\dot m$ & $\alpha$ & $\beta$   & $r_{\rm d}$ ($R_{\rm S}$) & $\dot m_{\rm cnd}$ ($\dot M_{\rm Edd}$)    & $\dot m
_{\rm cnd}$/ $\dot m$   & $T_{\rm eff,max} (\rm eV)$ & $\Gamma_{\rm 2-10 keV}$ \\
\hline
$10^{9}$ &  0.05  &0.3  &0.95     & 174  &  $2.13\times10^{-2}$    &42.6\%     &3.2    & 2.13\\
$10^{8}$ &  0.05  &0.3  &0.95     & 175  &  $2.13\times10^{-2}$    &42.6\%     &5.7    & 2.17\\
$10^{7}$ &  0.05  &0.3  &0.95     & 173  &  $2.12\times10^{-2}$    &42.4\%     &10.0   & 2.20\\
$10^{6}$ &  0.05  &0.3  &0.95     & 174  &  $2.13\times10^{-2}$    &42.6\%     &18.0   & 2.20\\
$10$     &  0.05  &0.3  &0.95     & 175  &  $2.13\times10^{-3}$    &42.6\%     &320.5  & 2.34\\
\hline            
$10^{9}$ &  0.1   &0.3  &0.95    & 446  &   $6.79\times10^{-2}$   &67.9\%    &4.0       & 2.24\\
$10^{8}$ &  0.1   &0.3  &0.95    & 444  &   $6.78\times10^{-2}$   &67.8\%    &7.1       & 2.29\\
$10^{7}$ &  0.1   &0.3  &0.95    & 450  &   $6.80\times10^{-2}$   &68.0\%    &12.7      & 2.34\\
$10^{6}$ &  0.1   &0.3  &0.95    & 450  &   $6.80\times10^{-2}$   &68.0\%    &22.6      & 2.41\\
$10$     &  0.1   &0.3  &0.95    & 452  &   $6.81\times10^{-2}$   &68.1\%    &402.5     & 2.68\\
\hline\hline
\end{tabular}
\\
\label{mass_effect}
\end{table*}


\subsection{The effect of $\alpha$}\label{secalpha}
\begin{figure*}
\includegraphics[width=88mm,height=65mm,angle=0.0]{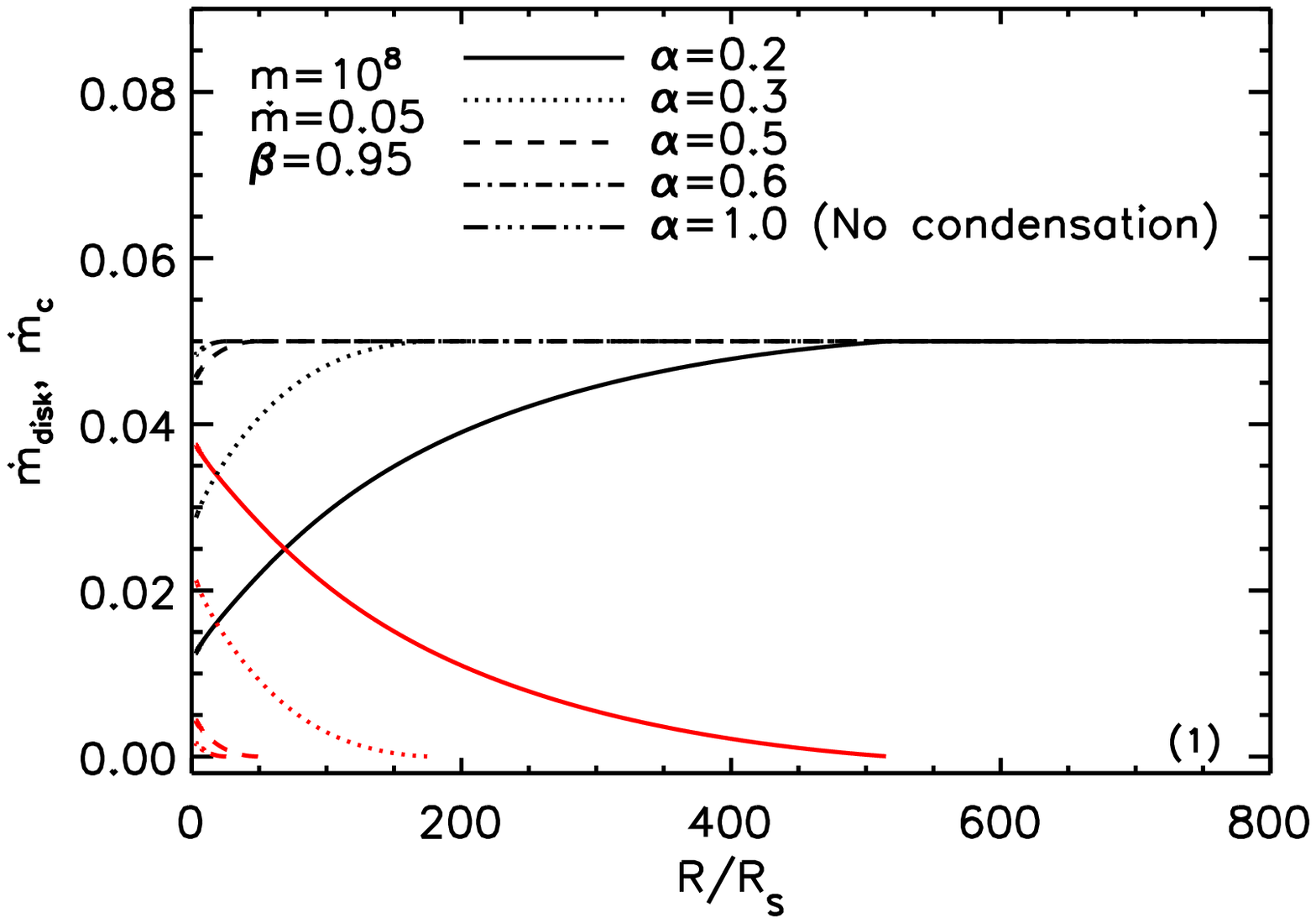}
\includegraphics[width=88mm,height=65mm,angle=0.0]{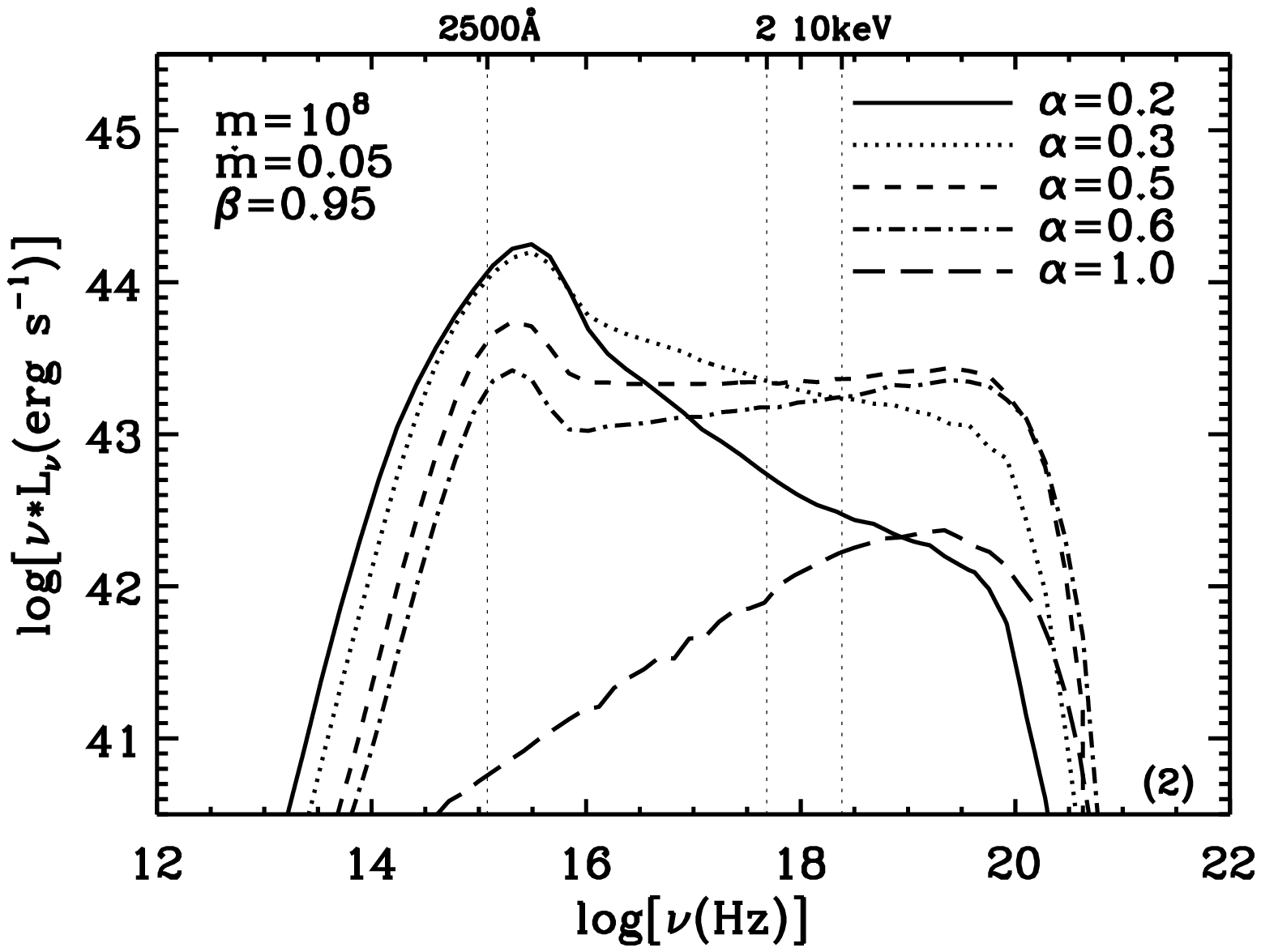}
\caption{\label{alpha}} Panel (1): mass accretion rate in the corona (black line) and mass
accretion rate in the disc (red line)  as a function of radius  for different  viscosity parameter $\alpha$.
Panel (2): corresponding  emergent spectra for different  viscosity parameter $\alpha$.
\end{figure*}

In the panel (1) of Fig. \ref{alpha}, we plot  the mass accretion rate in the corona (black line) and mass 
accretion rate in the disc (red line) as a function of radius 
with a fixed  initial mass accretion rate $\dot m=0.05$ for different $\alpha$.
It is clear that  the distribution of the mass accretion rate in the corona and the mass
accretion rate in the disc are strongly affected by $\alpha$. 
As shown in Table \ref{alpha_effect}, the corresponding  condensation radius  $r_{\rm d}$
and the condensation rate of the corona $\dot m_{\rm cnd}$ are also strongly affected by $\alpha$.
Specifically, for 
$\alpha=0.2$, the condensation radius is $r_{\rm d}=514$ and the condensation rate is 
$\dot m_{\rm cnd}=3.75\times 10^{-2}$. With the increase of $\alpha$, as examples, 
for $\alpha=0.3, 0.5, 0.6$, the condensation radii are $r_{\rm d}=175, 49, 26$ respectively, and
the condensation rates are $\dot m_{\rm cnd}=2.13\times 10^{-2}, 4.35\times 10^{-3}, 1.63\times 10^{-3}$
respectively. For $\alpha=1$, there is no condensation.
We plot the corresponding emergent spectra in the panel (2) of Fig. \ref{alpha}.
It is obvious that the emergent spectra are dramatically changed with changing $\alpha$,
which can be generally understood as follows. First, we should keep in mind that the self-similar 
solution of ADAF is employed  as the initial condition for describing the corona in our model. 
Then as we know, the luminosity of ADAF can be roughly expressed as, $L_{\rm ADAF} \propto \dot m^{2} \alpha^{-2}$,
which means that an increase of the value of $\alpha$ will decrease the radiation of the ADAF for 
a fixed mass accretion rate \citep{Mahadevan1997}. If the radiation decreases, physically as we can 
imagine that the gas in  ADAF is not easy to condense.
Meanwhile, generally the cooling rate of the accretion flow (not only ADAF) is inversely proportional 
to the radius from the black hole. So for a bigger value of $\alpha$, the hot corona/ADAF will begin to condense
at a relatively smaller radius from the black hole  where the cooling is strong enough so that 
the condensation can occur. Extremely, for $\alpha=1$, as we show in the panel (1) of Fig. \ref{alpha}, 
at any radius beyond the ISCO of the black hole, the condensation doesn't occur. 

From Table \ref{alpha_effect}, we can see, for $\alpha=0.2$, the ratio between the condensation rate
and the initial mass rate is $\dot m_{\rm cnd}/\dot m = 75.0\%$ 
and the hard X-ray index is $\Gamma_{\rm 2-10keV}=2.37$. With the increase of $\alpha$, 
as examples, for $\alpha=0.3, 0.5, 0.6$, $\dot m_{\rm cnd}/\dot m$ are $42.6\%, 8.7\%, 3.3\%$ respectively, and 
$\Gamma_{\rm 2-10keV}$ are $2.17, 1.96, 1.89$ respectively. For the extreme case $\alpha=1$, there is no
condensation and the hard X-ray index is $\Gamma_{\rm 2-10keV}=1.59$. 
As the numerical results show, with the increase of $\alpha$, $\dot m_{\rm cnd}/\dot m$ decreases
dramatically. It means that, with the increase of $\alpha$, more and more initially accreted coronal gas  will keep
the hot condition, only a smaller fraction of the hot corona condenses downward forming the cool disc, which
predicts a higher electron temperature in the corona, consequently resulting in a harder X-ray spectrum. 

\begin{table*}
\caption{Condensation and spectral features of the inner disc and corona for different viscosity parameter $\alpha$.
$r_{\rm d}$ is the condensation radius. $\dot m_{\rm cnd}$ is the integrated condensation rate.
$T_{\rm eff,max}$ is the maximum effective temperature of the inner disc. 
$\Gamma_{\rm 2-10 keV}$ is the hard X-ray photon index between 2 and 10keV.} 

\centering
\begin{tabular}{ccccccccc}
\hline\hline
$m$ & $\dot m$ & $\alpha$ & $\beta$   & $r_{\rm d}$ ($R_{\rm S}$)  & $\dot m_{\rm cnd}$ ($\dot M_{\rm Edd}$) & $\dot m
_{\rm cnd}$/ $\dot m$   & $T_{\rm eff,max} (\rm eV)$ & $\Gamma_{\rm 2-10 keV}$ \\
\hline
$10^{8}$ &  0.05  &0.2  &0.95     & 514  &  $3.75\times10^{-2}$    &75.0\%     &5.7    & 2.37\\
$10^{8}$ &  0.05  &0.3  &0.95     & 175  &  $2.13\times10^{-2}$    &42.6\%     &5.7    & 2.17\\
$10^{8}$ &  0.05  &0.5  &0.95     & 49   &  $4.35\times10^{-3}$    &8.7\%      &4.4    & 1.96\\
$10^{8}$ &  0.05  &0.6  &0.95     & 26   &  $1.63\times10^{-3}$    &3.3\%      &3.7    & 1.89\\
$10^{8}$ &  0.05  &1.0  &0.95     & -     &   -                      &   -        & -      & 1.59\\
\hline\hline
\end{tabular}
\\
\label{alpha_effect}
\end{table*}

\subsection{The effect of $\beta$}\label{secbeta}
\begin{figure*}
\includegraphics[width=88mm,height=65mm,angle=0.0]{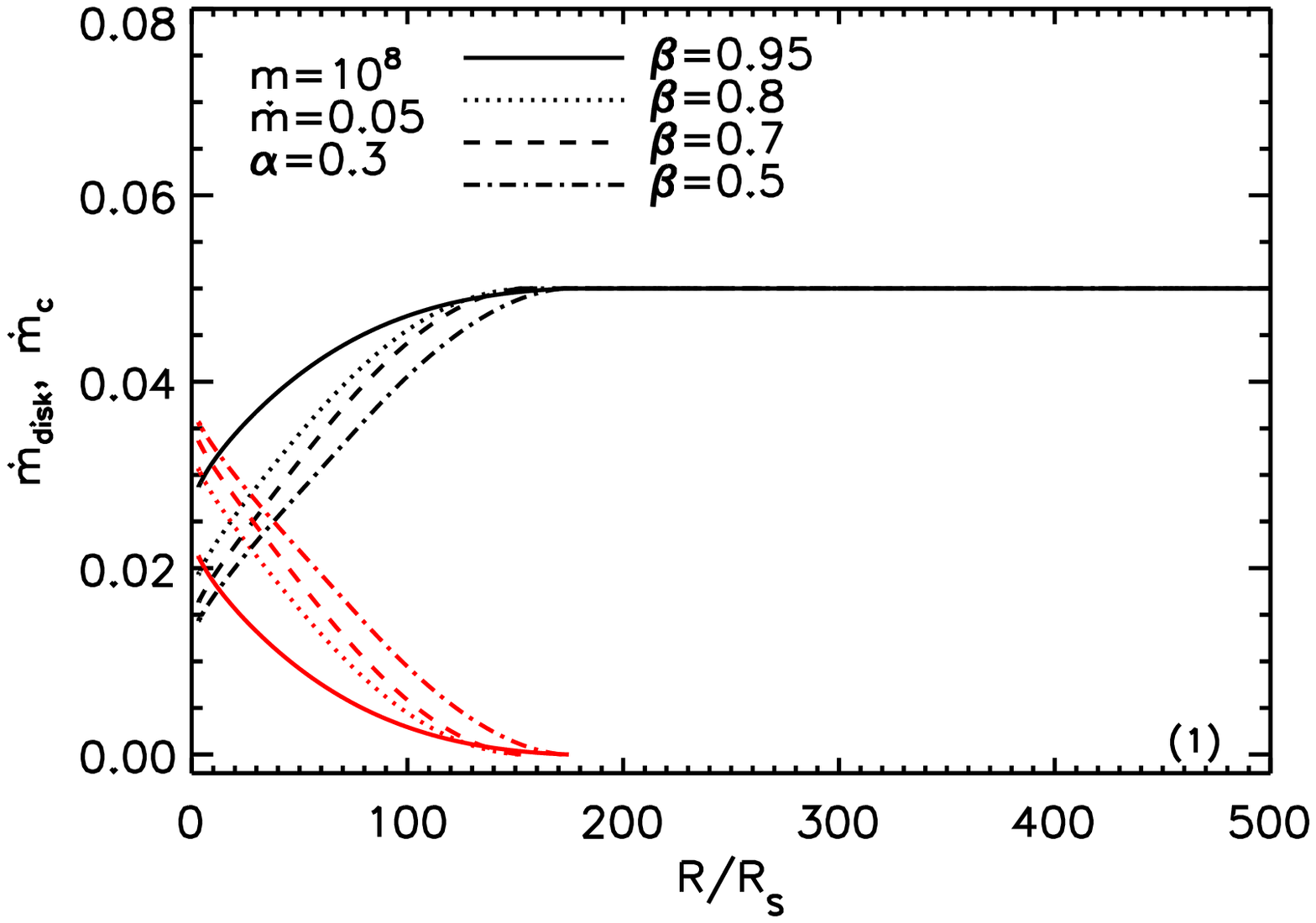}
\includegraphics[width=88mm,height=65mm,angle=0.0]{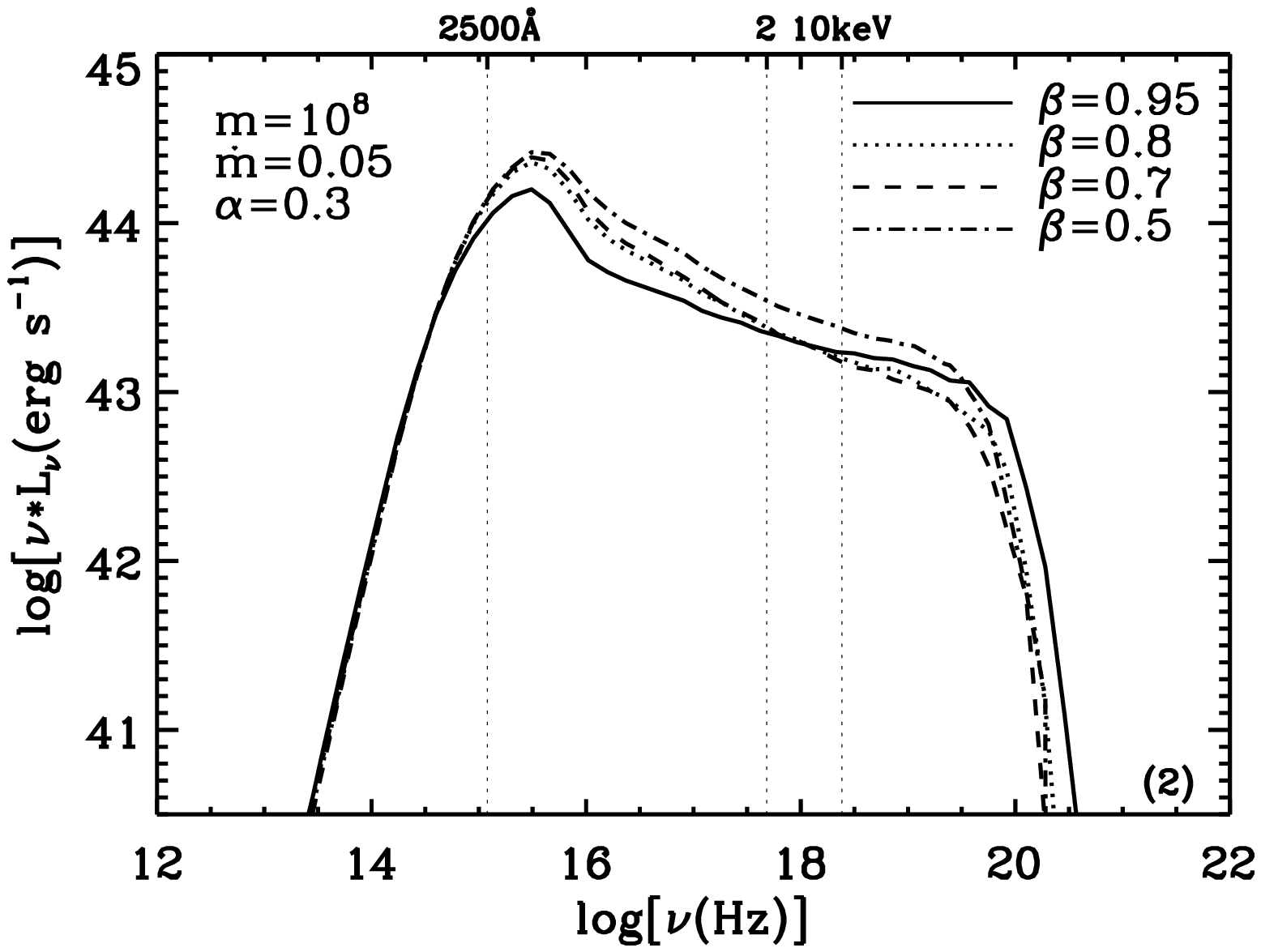}
\caption{\label{beta}}Panel (1): mass accretion rate in the corona (black line) and mass
accretion rate in the disc (red line)  as a function of radius  for different  magnetic parameter $\beta$.
Panel (2): corresponding  emergent spectra for different  magnetic parameter $\beta$.
\end{figure*}

In the panel (1) of Fig. \ref{beta}, we plot the mass accretion rate in the corona (black line) and mass 
accretion rate in the disc (red line) as a function of radius 
with initial mass accretion rate $\dot m=0.05$ for different $\beta$.
For $\beta=0.95$, the condensation radius  is $r_{\rm d}=175$, and the condensation rate 
is $\dot m_{\rm cnd}=2.13\times 10^{-2}$.
With the decrease of $\beta$, there is a slight change of the condensation radius and a slight 
increase of the condensation rate, i.e., for $\beta=0.8, 0.7$ and $0.5$,
$r_{\rm d}=151, 153$ and $171$ respectively, and 
$\dot m_{\rm cnd}=3.07\times 10^{-2}, 3.37\times 10^{-2}$ and $3.57\times 10^{-2}$ respectively.
The aforementioned results for the effects of $\beta$ on the condensation radius
and the condensation rate can be roughly understood as follows.
In both the current condensation model and the pure ADAF model, the magnetic field is responsible for the
synchrotron radiation of the corona/ADAF. In the pure ADAF model, the synchrotron radiation
is one of the most important soft-photon source for the Comptonization.
In our condensation model, when the mass accretion rate is greater than the critical mass accretion 
rate $\dot M_{\rm crit}$, the condensation occurs, forming an inner disc.
Generally, the energy density of the soft photons from the inner disc will dominate the soft photons 
from the synchrotron radiation of the corona itself for the Comptonization with the size of the 
inner disc greater than $\sim 5R_{\rm S}$. With the increase of the mass accretion rate, 
the soft photons from the disc will completely dominate the soft photons from the synchrotron radiation 
for the Comptonization. So a change of the magnetic parameter $\beta$, from 0.95 to 0.5, intrinsically
will not significantly change the model results, such as the condensation radius  and condensation rate.

We should also note that, indeed there is a slight increase of the
condensation rate with decreasing $\beta$, which we think can be understood from the hydrostatic equilibrium 
of the corona in the vertical direction. 
In the vertical direction, the gravity exerted on the corona is balanced by the total pressure gradient,  
$\partial p_{\rm tot}/\partial z \sim {p_{\rm tot}/H}$ (with $p_{\rm tot}=p_{\rm gas}+p_{\rm m}$, 
$H$ being the scaleheight of the corona).
A decrease of $\beta$ means an increase of the magnetic filed, which leads to an increase of the magnetic pressure.
In order to keep the hydrostatic equilibrium of the corona in the vertical direction, 
if the total pressure $p_{\rm tot}$ and the scaleheight $H$ do not change, an increase of the magnetic pressure  
means the gas pressure $p_{\rm gas}$ decreases. 
According to the equation of state of the corona, the gas pressure is 
$p_{\rm gas} \varpropto \rho (T_{\rm i}+ T_{\rm e}) \sim \rho T_{\rm i}$ (with $T_{\rm i}$ being 
the ion temperature, $T_{\rm e}$ being  the electron temperature and $\rho$ being the density of the corona respectively).
Generally, the ion temperature $T_{\rm i}$ in the corona changes very slightly, 
which is close to the virial temperature. So when the gas pressure  $p_{\rm gas}$ decreases, 
the density $\rho$ decreases. The mass accretion rate in the corona can be expressed as,
$\dot M_{\rm cor}=4\pi R v_{\rm R} \rho H \sim 4\pi R \cdot  \alpha c_{\rm s} \big({H\over R}\big) \cdot \rho H
\varpropto c_{s}^{2} \rho H \varpropto T_{i} \rho H$. As our analysis above, when $p_{\rm gas}$ decreases,
$\rho$ decreases, meanwhile, $T_{\rm i}$ and $H$ nearly does not change, so $\dot M_{\rm cor}$ decreases, which 
means that the condensation rate increases. We address that the analysis above is very simple, 
one can refer to Table \ref{beta_effect} for the detailed numerical results.

In the panel (2) of Fig. \ref{beta}, we plot the corresponding emergent spectra with 
$\dot m=0.05$ for different $\beta$. As we can see, there is no obvious change of the hard X-ray spectra,
e.g., for $\beta=0.95, 0.8, 0.7$ and $0.5$, the hard X-ray indices between 2-10keV are 
$\Gamma_{\rm 2-10keV}=2.17, 2.26, 2.28$ and $2.22$ respectively.

\begin{table*}
\caption{Condensation and spectral features of the inner disc and corona for different magnetic parameter $\beta$.
$r_{\rm d}$ is the condensation radius. $\dot m_{\rm cnd}$ is the integrated condensation rate.
$T_{\rm eff,max}$ is the maximum effective temperature of the inner disc. 
$\Gamma_{\rm 2-10 keV}$ is the hard X-ray photon index between 2 and 10keV.}

\centering
\begin{tabular}{ccccccccc}
\hline\hline
$m$ & $\dot m$ & $\alpha$ & $\beta$  & $r_{\rm d}$ ($R_{\rm S}$) & $\dot m_{\rm cnd}$ ($\dot M_{\rm Edd}$) & $\dot m
_{\rm cnd}$/ $\dot m$   & $T_{\rm eff,max} (\rm eV)$ & $\Gamma_{\rm 2-10 keV}$ \\
\hline
$10^{8}$ &  0.05  &0.3  &0.95     & 175   &  $2.13\times10^{-2}$    &42.6\%     &5.7    & 2.17\\
$10^{8}$ &  0.05  &0.3  &0.8      & 151   &  $3.07\times10^{-2}$    &61.4\%     &6.4    & 2.26\\
$10^{8}$ &  0.05  &0.3  &0.7      & 153   &  $3.37\times10^{-2}$    &67.4\%     &6.6    & 2.28\\
$10^{8}$ &  0.05  &0.3  &0.5      & 171   &  $3.57\times10^{-2}$    &71.4\%     &6.9    & 2.22\\
\hline\hline
\end{tabular}
\\
\label{beta_effect}
\end{table*}

\subsection{Comparison with observations for the $\Gamma_{\rm 2-10keV}-L_{\rm bol}/L_{\rm Edd}$
correlation in luminous AGNs---the effect of $\alpha$}\label{observation}
\begin{figure*}
\includegraphics[width=95mm,height=70mm,angle=0.0]{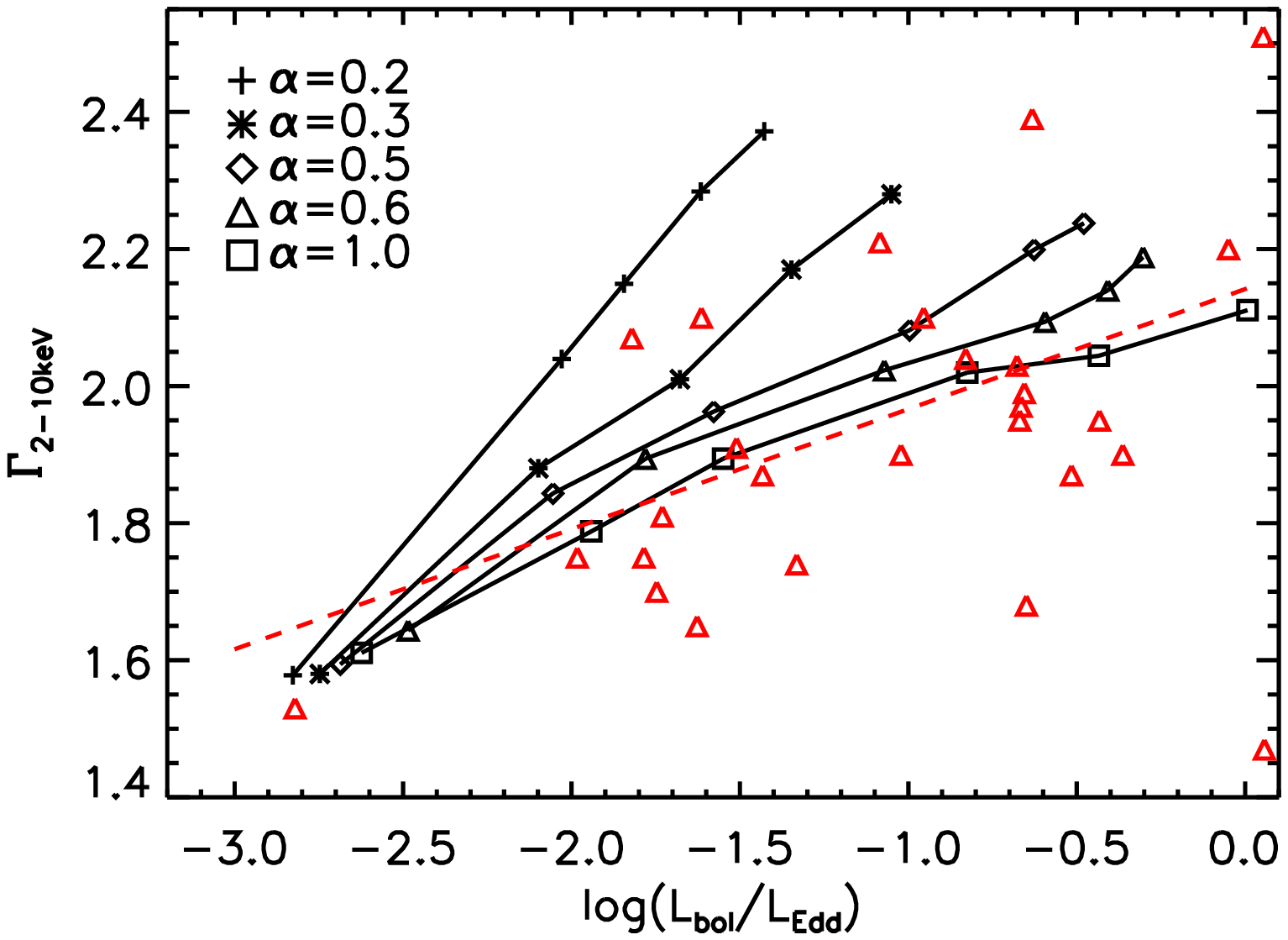}
\caption{\label{gama}} Hard X-ray photon index $\Gamma_{\rm 2-10keV}$ as a function of Eddington ratio 
$L_{\rm bol}/L_{\rm Edd}$. The red-triangle is a sample composed of 29 luminous AGNs from \citet{Vasudevan2009}. 
The red-dashed line refers to the best-fitting linear regression of the observational data.
The symbol '+', $\ast$, $\lozenge$, $\triangle$, and $\square$ refer to the theoretical results 
for $\alpha=0.2, 0.3, 0.5, 0.6$ and $1.0$ respectively.
In all the calculations, $m=10^8$ and $\beta=0.95$ are adopted.
\end{figure*}

As shown in Section \ref{secmass},  \ref{secalpha} and  \ref{secbeta}, we studied the effects 
of the black hole mass $M$, viscosity parameter $\alpha$ and the magnetic parameter $\beta$ on the
structure of the disc and the corona, as well as the 
corresponding emergent spectra respectively. Specifically, it is found that the effects of $M$
and $\beta$ on the hard X-ray photon index $\Gamma_{\rm 2-10keV}$  are very little, which nearly can be neglected.
In the current paper, we will focus on the effect of $\alpha$ on $\Gamma_{\rm 2-10keV}-L_{\rm bol}/L_{\rm Edd}$ correlation.  
In the calculation, we fix black hole mass $m=10^8$, magnetic parameter $\beta=0.95$   
to calculate the emergent spectra with $\dot m$ by assuming a value of $\alpha$. Specifically,  
for $\alpha=0.2$, $\dot m=0.05, 0.03, 0.02, 0.015$ and $0.009$ are adopted to calculate the emergent spectra respectively .
For $\alpha=0.3$, $\dot m=0.1, 0.05, 0.03, 0.02$ and $ 0.015$ are adopted respectively.
For $\alpha=0.5$, $\dot m=0.3, 0.2, 0.1,  0.05, 0.035$ and $0.03$ are adopted respectively.
For $\alpha=0.6$, $\dot m=0.4, 0.3, 0.2, 0.1, 0.05$ and $0.04$ are adopted respectively.
For $\alpha=1$, $\dot m=0.6, 0.3, 0.2, 0.1, 0.08$ and $0.07$ are adopted respectively.
The hard X-ray photon index $\Gamma_{\rm 2-10keV}$ can be directly got from the theoretical X-ray spectra.
The bolometric luminosity $L_{\rm bol}$ can be got by integrating the emergent spectra.
We plot the hard X-ray photon index  $\Gamma_{\rm 2-10keV}$ as a function of $L_{\rm bol}/L_{\rm Edd}$  
in Fig. \ref{gama} for different $\alpha$.
The symbol '+', $\ast$, $\lozenge$, $\triangle$, and $\square$ refer to the theoretical results 
for $\alpha=0.2, 0.3, 0.5, 0.6$ and $1.0$ respectively. 
As shown in Fig. \ref{gama}, it is clear that there is positive correlation between 
$\Gamma_{\rm 2-10keV}$ and $L_{\rm bol}/L_{\rm Edd}$ by assuming a
fixed value of $\alpha$, which can be roughly understood as our general picture of the condensation model.
With an increase of the initial mass accretion rate, both the condensation radius $r_{\rm d}$ and 
the condensation rate $\dot m_{\rm cnd}$ increases,  then more photons in the disc to be scattered 
in the corona, which will lead to a lower electron temperature in the corona, consequently  predicting a 
softer X-ray spectrum. 

We collected a sample composed of 29 luminous AGNs with well constrained X-ray spectra and Eddington ratios
for comparisons \citep{Vasudevan2009}.
The black hole masses of the sources in the sample are estimated by the  
reverberation mapping (RM) method, which is believed to be a proven and powerful technique for the
estimation of the black hole mass in AGNs \citep{Peterson2004}. The bolometric luminosity of the sources in the
sample are calculated by integrating the spectral energy distributions (SEDs) of the sources with 
simultaneous X-ray/optical/UV data of the \emph{XMM}-\emph{Newton}
European Photon Imaging Camera (EPIC-pn) and Optical Monitor (OM) archives 
\citep{Vasudevan2009}. 
The X-ray spectra are fitted by the \textsc{bknpower} model, and the model of    
\textsc{zwabs} was allowed for treating the absorptions \citep{Vasudevan2009}.
One can refer to the red-triangle in Fig. \ref{gama}
for the observational data between $\Gamma_{\rm 2-10keV}$ and $L_{\rm bol}/L_{\rm Edd}$.
The red-dashed line in Fig. \ref{gama}
refers to the best-fitting linear regression of the observational data, which
can be expressed as,
\begin{eqnarray}\label{pfit}
\Gamma_{\rm 2-10keV}=2.1+0.18\times {\rm log} ({L_{\rm bol}/L_{\rm Edd}})
\end{eqnarray}
It is obvious that the theoretical $\Gamma_{\rm 2-10keV}-L_{\rm bol}/L_{\rm Edd}$ correlation derived by assuming a
bigger $\alpha$, i.e., $\alpha \sim 1$ is more close  to the observations.

\section{Discussions}
\subsection{On black hole mass}
As has been shown in  \ref{secmass}, the effects of black hole mass on the distribution of the mass accretion 
rate in the corona and mass accretion rate in the disc, the electron temperature in the corona, the Compton
scattering optical depth of the corona can nearly be neglected. Meanwhile, since the effective temperature of the
disc for a higher mass black hole is intrinsically lower than that of the case for a lower black hole mass 
\citep{Mitsuda1984,Makishima1986}, the emergent spectrum nearly shifts rightward horizontally 
with decreasing black hole mass, as shown in the panel (1) and panel (2) of Fig. \ref{mass2}.
We note that, in the current paper, the thermal conduction between 
the disc and corona in the vertical direction is considered , which finally results in  the 
formation of the inner disc. Because of the feedback between the corona and the disc, the 
distribution of the electron temperature in the corona and the Compton scattering optical depth of the corona
as a function of radius are different from the results predicted by the pure ADAF \citep{Narayan1995}.
However, since the corona in our model intrinsically is optically thin,  
the results predicted by our model for supermassive black holes in AGNs theoretically 
can be scaled down to stellar mass black holes, which is similar to the case of pure ADAF solution 
applied to both supermassive black holes and stellar mass black hole \citep[][for review]{Yuan2014}.
A detailed study of the application of the condensation of the corona to high
mass black hole X-ray binaries (such as Cyg X-1) will be in the next paper, which is believed to 
be fuelled by the wind loss from the bright companion star.  

\subsection{On $\alpha$ and $\beta$}
The viscosity parameter $\alpha$ and magnetic parameter $\beta$ are two very important parameters in 
both our current model of the condensation of the corona and the pure ADAF model.  
A great of very important progresses have been done for investigating the value of $\alpha$ and $\beta$
or the relationship between $\alpha$ and $\beta$ for different types of the accretion flows or for the 
accretion flow with different  magnetic structure. 
\citep[e.g.][]{Pessah2007,Blackman2008,Guan2009,Hawley2011,McKinney2012,Bai2013}.
Since we employed the pure ADAF solution as the initial condition for  describing the property of the 
corona in the current model, we will focus on the value of $\alpha$ and $\beta$ 
in the framework of the ADAF solution. 

The viscosity is one of the most important processes to be considered in the accretion physics,
which basically controls how  the angular momentum is transported and accreted gas is heated.
\citet[][]{Shakura1973} proposed a very simple description for the viscosity, which was expressed 
as follows, 
\begin{eqnarray}\label{pfit}
\nu=\alpha c_{\rm s} H,
\end{eqnarray}
where $\nu$ refers to the kinematic viscosity, $c_{\rm s}$ refers to the sound speed, and $H$ refers
to the thickness of the disc. The value of $\alpha$ was expected 
$\lesssim 1$ \citep[e.g.][for detailed discussions]{Frank2002}.
Such a dimensionless  description for the viscosity was widely applied to nearly all types of the semi-analytical
study of the accretion flows around a black hole,
such as the slim disc with a higher mass accretion rate by \citet[][]{Abramowicz1988}, the ADAF with
a lower mass accretion rate by \citet[][]{Narayan1994,Narayan1995}.
Currently, it is widely believed that the  physical mechanism of the viscosity 
for the angular momentum transport in the ionized accretion flows is 
the magnetorotational instability (MRI) \citep[][]{Balbus1991,Balbus1998}.
So far a great of MHD numerical simulations have been done for the viscosity parameter $\alpha$, however,
it is still uncertain, depending strongly on how the numerical simulations are considered \citep[][for review]{Yuan2014}.
Previously, in some numerical simulations, the author generally found a lower value of $\alpha$, such as, 
$\alpha<0.01$ by \citet{Stone1996}, $\alpha\simeq 0.016$ by \citet{Hirose2006},
and $\alpha \sim 0.003-0.01$ by \citet{Hawley2011}.
Reversely,  some numerical simulations found a relatively bigger value of $\alpha$ , sometimes even 
greater than unity \citep{Machida2000,Bai2013}. 

Observationally, the value of $\alpha$ is also diverse.
\citet{Starling2004} estimated  the value of $\alpha$ by measuring the optical variability  
of AGNs from months to years.
By analysing a two-folding time-scale observational data in the optical band, they suggested that 
$\alpha$ in the disc is  in the range of $0.01-0.03$ for $0.1\lesssim L_{\rm bol}/L_{\rm Edd} \lesssim 1$.
\citet{Starling2004} also reminded that such a value of $\alpha$ is  only a lower limit since the data sampling 
may miss the shorter time-scales.
\citet{King2007} summarized the value of $\alpha$ in different kinds of objects, including Dwarf nova outburst,
X-ray transients outburst, AGNs, proto-stellar accretion discs, FU Orionis outbursts. 
They suggested that the value of $\alpha$  is in the range of $0.1-0.4$.
Based on the disk-evaporation model, \citet{Qiao2009} studied the effect of $\alpha$ on the transition luminosity
$L_{\rm tr}$ between the low/hard spectral state and high/soft 
spectral state in black hole X-ray binaries. Specifically, they found that $L_{\rm tr}/L_{\rm Edd} \propto \alpha^{2.34}$. 
By comparing with a sample of black hole X-ray binaries with well constrained transitional luminosities
(ranging from $0.0069-015 L_{\rm Edd}$), the authors suggested that the value of $\alpha$ is in the range of $0.1-0.6$.

Although the ADAF solution was originally proposed for the low-luminosity accreting black holes, it was suggested that
ADAF can also be employed to describe the high-luminosity accreting black holes, providing a larger 
$\alpha$, e.g., $\alpha \sim 1$ was adopted in \citet[][]{Narayan1996}. As we know, theoretically there is a critical 
mass accretion rate $\dot M_{\rm crit}^{'}$, i.e., $\dot M_{\rm crit}^{'}\propto \alpha^2$, above which the ADAF solution
can not exist. It is clear that the critical mass accretion rate $\dot M_{\rm crit}^{'}$ 
increases with increasing $\alpha$. So the ADAF solution can be applied to the black hole with a higher luminosity.  
\citet{Xie2016} employed the coupled  ADAF-jet model to explain the radio/X-ray correlation 
of $L_{\rm R}\propto L_{\rm x}^{\sim 0.6}$ in  black hole X-ray transients GX339-4 and 
V404 Cyg for $L/L_{\rm Edd} \gtrsim 10^{-3}$, in which the X-ray emission is dominated by ADAF 
with a relatively bigger value of $\alpha=0.6$ adopted  and the radio emission is dominated by jet.

In many numerical simulations, it is found that there is a tight correlation between $\alpha$ and $\beta^{'}$,
i.e., $\alpha \beta^{'} \sim 0.5$ (with $\beta^{'}=p_{\rm gas}/p_{\rm m}$, different from the 
previous definition for the magnetic parameter as,  $\beta=p_{\rm gas}/{(p_{\rm gas}+p_{\rm m})}$)
\citep[e.g.][]{Blackman2008,Hawley2011,Sorathia2012}.
For example, \citet{Hawley2011} found that $\beta^{'}$ is in the range of $\sim 10-200$,
corresponding to a lower value of $\alpha$, i.e., $\alpha \sim 0.01-0.003$.
However, we should note that, as far as ADAF solution, MHD simulations usually give $\beta^{'} \simeq 10$ 
(corresponding to $\beta \simeq 0.91$), which means a relatively weak magnetic field in the ADAF solution. 
\citet{Qiaoetal2013} estimated the value of $\beta$ in the ADAF solution by comparing with a sample composed
10 low-luminosity AGNs with well observed SEDs. They also found  that a weaker magnetic field, i.e., 
$\beta \simeq 0.95$ is required to fit the observed SEDs. In the current paper, as has been 
shown in section \ref{secbeta}, the effects of $\beta$ on the X-ray spectra are  very little,
so we simply fix $\beta=0.95$ in section \ref{observation} to compare 
with observations.

\section{Conclusions}
In this paper, we systemically studied the effects of the black hole $M$, viscosity parameter $\alpha$,
and the magnetic parameter $\beta$ on the structure of the disc and corona, as well as the emergent spectra  
based on the model of the condensation of the corona. It is found that the slope of the hard X-ray spectra
nearly does not change with changing $M$ or $\beta$. Meanwhile, it is found that the 
geometry of the accretion flow, the structure of the disc and the corona, as well as the corresponding emergent spectrum 
completely changes with changing $\alpha$.
By comparing with a sample composed of 29 luminous AGNs with well constrained X-ray spectra and Eddington ratios 
$L_{\rm bol}/L_{\rm Edd}$, it is found that the observed  
correlation between $\Gamma_{\rm 2-10keV}$  and $L_{\rm bol}/L_{\rm Edd}$  can be well matched by taking $\alpha \sim 1$
in our model, which has been proposed to explain the spectral features of luminous accreting 
black holes in the framework of ADAF solution, e.g., \citet{Narayan1996}. 


\section*{Acknowledgments}
We thank Professor R.E. Taam from ASIAA for very useful discussions. 
This work is supported by the National Natural Science Foundation of
China (Grants 11773037 and 11673026), the gravitational wave pilot B (Grants No. XDB23040100), 
and the National Program on Key Research and Development Project (Grant No. 2016YFA0400804).

\bibliographystyle{mnras}
\bibliography{qiaoel}



\bsp	
\label{lastpage}
\end{document}